\documentclass[8pt,reqno]{amsart}
\usepackage[applemac]{inputenc}
\usepackage[english]{babel}
\usepackage{latexsym}
\usepackage{amssymb}
\usepackage{amsmath}
\usepackage{amsthm}
\usepackage{nccmath}
\usepackage{amsfonts}
\usepackage{amssymb} 

\numberwithin{equation}{section}

\newcommand{\be}{\begin{equation}}
\newcommand{\ee}{\end{equation}}
\newcommand{\bes}{\begin{equation*}}
\newcommand{\ees}{\end{equation*}}

\newcommand{\eqn}{\begin{eqnarray}}
\newcommand{\feqn}{\end{eqnarray}}
\newcommand{\eqnn}{\begin{eqnarray*}}
\newcommand{\feqnn}{\end{eqnarray*}}

\newcommand{\smaq}{\left[ \begin{smallmatrix}}
\newcommand{\smat}{\left( \begin{smallmatrix}}
\newcommand{\smcq}{\end{smallmatrix}\right]}
\newcommand{\smct}{\end{smallmatrix}\right)}
\newcommand{\smag}{\left \{ \begin{smallmatrix}}
\newcommand{\smcg}{\end{smallmatrix}\right \}}

\DeclareMathOperator{\SO}{SO}

\DeclareMathOperator{\Vol}{Vol}

\begin{document}

\title[A simple $E_8$ construction]{A simple \boldmath{$E_8$} construction}

\author{Sergio L. Cacciatori}
\author{Francesco Dalla Piazza}
\address{
Dipartimento di Scienze e Alta Tecnologia, Universit\`a degli Studi
dell'Insubria, Via Valleggio 11, 22100 Como, Italy, and INFN, via
Celoria 16, 20133 Milano, Italy}
\email{sergio.cacciatori@uninsubria.it}
\email{f.dallapiazza@uninsubria.it}

\author{Antonio Scotti}
\address{Dipartimento di Fisica,
Universit\`a degli Studi di Milano,  Via Celoria 16, 20133 Milano, Italy}
\email{ascotti@mindspring.com}


\begin{abstract}
In this short letter we conclude our program, started in \cite{Cacciatori:2005yb}, of building up explicit
generalized Euler angle parameterizations for all exceptional compact Lie groups. In this last step we solve the problem for $E_8$.
\end{abstract}

\maketitle

\section{Introduction}
In \cite{Cacciatori:2005yb}, we started a project having the aim of providing an explicit realization of all compact (simply connected)
exceptional Lie groups, including an explicit determination of the range of parameters in order to cover the group exactly one time. The problem
of specifying the range is indeed a complication in several numerical simulations involving global properties of the group and is really time
consuming if it has to  be treated numerically. To this end, we have introduced the generalized Euler parametrization, which has the advantage
of characterizing the complete ranges in a very simple form (see \cite{rev} for a review). We have done this for all the exceptional Lie groups but $E_8$ in
\cite{G2}, \cite{F4}, \cite{E6}, and \cite{e7}. Here we complete our program by providing a generalized Euler parametrization of $E_8$ starting
from its maximal compact subgroup S$_s$(16)$\equiv$Spin(16)/$\mathbb{Z}_2$. We will employ a general method
developed in \cite{dyson}, and refer, for the details proving our assertions, to that paper.

It is worth to mention that, beyond the applications in string theories, GUT theories and theories of everything \cite{L,LSS}, the exceptional Lie group $E_8$ finds applications also in more phenomenological areas of physics like, for example, the quasi-one-dimensional Ising ferromagnet ${\rm CoNb_2O_6}$ described in \cite{CK}.

\section{The algebra}
A very simple construction of the $E_8$ Lie algebra is given in \cite{adams}, chapter 6 (see also chapter 7). The compact real form is obtained by starting
from the Lie algebra $L:=$Lie(Spin(16)) (or, equivalently by its adjoint irrep. $\mathbf {120}$) and its real irreducible representation
$\Delta_+\equiv {\mathbf {128}}$. Let us review how this construction works concretely. \\
The Spin(16) subalgebra can be thought of as generated by elements $J_{ij}=-J_{ji}$, $1\leq i<j\leq 16$ with commutators
\be\label{spin16}
[J_{ij},J_{kl}]=\delta_{jk}J_{il}-\delta_{jl}J_{ik}-\delta_{ik}J_{jl}+\delta_{il}J_{jk} \equiv \sum_{m<n} {C_{ij,kl}}^{mn} J_{mn}.
\ee
The support space for $\Delta_+$ has generators $Q_\alpha$, $\alpha=1,\ldots, 128$, transforming as a Majorana-Weyl spinors of Spin(16).\footnote{These
can be easily obtained by the very useful Mathematica package gamma.m by Jeremy Michelson, available on http://www.physics.ohio-state.edu/~jeremy/}
Let $\gamma_j$, $j=1,\ldots, 16$ be the $128 \times 128$ gamma matrices generating the corresponding Clifford algebra. Thus, the representation
$\Delta_+$ has generators
\begin{eqnarray}
\Delta_{ij}:=\Delta_+(J_{ij})=\frac 14 [\gamma_i,\gamma_j].
\end{eqnarray}
The adjoint representation $\mathbf {248}$ of the $E_8$ algebra is then completely specified by adding to the commutators rules \eqref{spin16}
the remaining commutators
\begin{align*}
[J_{ij},Q_\alpha]&=\frac 14 \sum_\beta [\gamma_i,\gamma_j]_{\alpha\beta}Q_\beta \\
[Q_\alpha,Q_\beta]&=\frac 18 \sum_{i,j}[\gamma^i,\gamma^j]_{\alpha\beta}J_{ij}.
\end{align*}
In this way we get the adjoint representation $\rho$ with generators $\rho(J_{ij}), \rho(Q_\alpha)$, $1\leq i<j\leq 16$, $\alpha=1,\ldots, 128$ given by
\begin{eqnarray}
&& M_{ij} :=\rho(J_{ij}) =\begin{pmatrix}
(M_{ij})_{kl,mn} & 0 \\
0 & (M_{ij})_{\alpha\beta}
\end{pmatrix}=
\begin{pmatrix}
{C_{ij,kl}}^{mn} & 0 \\
0 & (\Delta_{ij})_{\alpha \beta}
\end{pmatrix},
\qquad 1\leq i<j\leq 16, \\
&& M_{\alpha} :=\rho(Q_\alpha) =\begin{pmatrix}
0 & (M_{\alpha})_{kl,\beta} \\
(M_{\alpha})_{\gamma,mn}
\end{pmatrix}=
\begin{pmatrix}
0 & 4 (\Delta^{kl})_{\alpha\beta} \\
-4(\Delta^{mn})_{\alpha \gamma} & 0
\end{pmatrix},
\qquad \alpha=1,\ldots, 128.
\end{eqnarray}
A realization of this matrices is given by the Mathematica program on http://www.dfm.uninsubria.it/E8/, useful for the explicit computations. In particular, using that program
one can check for example that a possible choice of a Cartan subalgebra is given by the matrices $\rho(Q_\alpha)$ with
$\alpha=1, 8, 26, 31, 43, 46, 52, 53$. We rename these matrices $C_a$, $a=1,\ldots, 8$.


\section{The group}\label{group}
We can now construct the group $E_8$ by specializing the Euler parametrization and specifying the range of parameters. As shown in \cite{adams}, chapter 7, the
maximal compact subgroup corresponding to the SO(16) subalgebra is a Spin(16)$/\mathbb{Z}_2$ group. More specifically (see \cite{IY}) there are three possible
non isomorphic quotients of this kind, which are the SO group SO(16) and two ``semispin'' groups denoted S$_{s}$(16) and S$'_{s}$(16). The quotient
realized here is the semispin group S:=S$_{s}$(16) \cite{IY}. \\
Let $S[x_1, \ldots, x_{120}]$ any parametrization of S, for example an Euler parametrization. This is standard and we will not discuss it any further here.
With this at hand, a parametrization of $E_8$ is given by
\begin{eqnarray}
E_8[x_1, \ldots, x_{120}; y_1, \ldots, y_8; z_1, \ldots, z_{120}]=
(S[x_1, \ldots, x_{120}]/\mathbb{Z}_2^8) e^{\sum_{a=1}^8 y^a C_a} S[z_1, \ldots, z_{120}].
\end{eqnarray}
Here, the range for the $z$ coordinates must be chosen such to cover S exactly one times (up to a subset of zero measure). The action of the $\mathbb{Z}_2^8$
finite group in the left factor eliminates redundances and is realized simply by restricting the range of the parameters $x$ (w.r.t. the ranges of the $z_i$).
For example, in the generalized Euler construction of S, all parameters can be chosen periodic, but only 8 parameters cover the whole period in the
definition of the ranges. In this case, the action of the finite group just reduces these ranges to half the period. \\
In this way it remains to specify the range for the $y$ coordinates. By choosing $y_a$ in a period one gets a quite large redundance, so that a strategy
to reduce the ranges is necessary. A general solution of this problem is given in \cite{dyson}, and works as follows. Let $\alpha^1, \ldots, \alpha^8$ be
any given choice of simple roots wit components $(\alpha^i_1,\ldots, \alpha^i_8)$ w.r.t. the basis $C_1,\ldots, C_8$ of the Cartan subalgebra. Moreover,
let $\alpha^L =\sum_{a=1}^8 n_a \alpha^a$ the longest root. For $E_8$, $(n_1,\ldots, n_8)=(2,3,4,6,5,4,3,2)$. Thus, the (almost everywhere) injective
range for the parameters $y^a$ is
\begin{eqnarray}
&& 0\leq \sum_{a=1}^8 \alpha^i_a y^a <\pi, \qquad\ i=1,\ldots, 8, \\
&& 0\leq \sum_{a=1}^8 \alpha^L_a y^a <\pi.
\end{eqnarray}
In our case we get that a possible choice of simple roots is
\begin{eqnarray*}
&& \alpha^1=\frac 12 (1,-1,-1,-1,-1,-1,-1,1), \\
&& \alpha^2=(1,1,0,0,0,0,0,0), \\
&& \alpha^3=(0,-1,1,0,0,0,0,0), \\
&& \alpha^2=(0,0,-1,1,0,0,0,0), \\
&& \alpha^2=(0,0,0,-1,1,0,0,0), \\
&& \alpha^2=(0,0,0,0,-1,1,0,0), \\
&& \alpha^2=(0,0,0,0,0,-1,1,0), \\
&& \alpha^2=(0,0,0,0,0,0,-1,1),
\end{eqnarray*}
so that the longest root is $\alpha^L=(0,0,0,0,0,0,1,1)$. Then, the range of the parameters is specified by
\begin{align*}
0&\leq\frac{1}{2}(y^1-y^2-y^3-y^4-y^5-y^6-y^7+y^8)<\pi \\
0&\leq y^1+y^2<\pi \\
0&\leq y^2-y^1<\pi \\
0&\leq y^3-y^2<\pi \\
0&\leq y^4-y^3<\pi \\
0&\leq y^5-y^4<\pi \\
0&\leq y^6-y^5<\pi \\
0&\leq y^7-y^6<\pi \\
0&\leq y^7+y^8<\pi,
\end{align*}
or equivalently
\begin{align*}
0 &\leq y^1 < \frac{\pi}{6} \\
y^1 &\leq y^2 < \frac{\pi + y^1}{7}\\
y^2 &\leq y^3 < \frac 16 (\pi + y^1 - y^2) \\
y^3 &\leq y^4 < \frac{1}{5} (\pi + y^1 - y^2 - y^3) \\
y^4 &\leq y^5 < \frac{1}{4} (\pi + y^1 - y^2 - y^3 - y^4) \\
y^5 &\leq y^6 < \frac 13 (\pi + y^1 - y^2 - y^3 - y^4 - y^5) \\
y^6 &\leq y^7 < \frac 12 (\pi + y^1 - y^2 - y^3 - y^4 - y^5 - y^6) \\
-y^1& + y^2 + y^3 + y^4 + y^5 + y^6 + y^7 \leq y^8 < \pi - y^7.
\end{align*}

\end{document}